\documentclass[twocolumn,noshowpacs,preprintnumbers,amsmath,amssymb]{revtex4}

\usepackage{verbatim}

\usepackage{graphicx}
\usepackage{dcolumn}
\usepackage{bm}
\usepackage{wrapfig}

\begin{document}

\title{Iron nanoparticle formation in a metal-organic matrix: from ripening to gluttony}
\author{Christian Klinke}
\email{klinke@chemie.uni-hamburg.de}
\affiliation{Institute of Physical Chemistry, University of Hamburg, D - 20146 Hamburg, Germany}

\author{Klaus Kern}
\affiliation{Institut de Physique des Nanostructures, Ecole Polytechnique F\'ed\'erale de Lausanne,\\
CH - 1015 Lausanne, Switzerland \\ \textnormal{and} \\ Max-Planck-Institut f\"ur Festk\"orperforschung, D - 70569 Stuttgart, Germany \\}

\begin{abstract} 

A simple method for the fabrication of metal nanoparticles is introduced. Heating metal-organic crystals in vacuum results in the formation of well defined metal particles embedded in a carbon matrix. The method is demonstrated for iron-phthalocyanine. At 500$^{\circ}$C homogeneously distributed iron nanoparticles with a reasonably narrow size distribution form by nucleation and ripening. After this initial phase the formation kinetics changes drastically. The particles move in the matrix to incorporate material. The "gluttony" phase shows astonishing similarities with the search for nutrition of living micro-organisms. Particle formation, ripening and gluttony are followed in-situ by transmission electron microscopy. 

\end{abstract}

\maketitle

Nanometer-sized particles of magnetic metals are attracting substantial interest in basic and applied research since many years~\cite{C01,C02,C03,C04}. This activity is mainly driven by the envisaged application of such nanoparticles in high-density data storage~\cite{C05}, in magnetic resonance imaging~\cite{C06,C07}, and as catalysts~\cite{C08,C09}. In many cases they show improved or even new properties due to their size in the nanometer range~\cite{C10}. State-of-the-art production methods of metal nanoparticles include arc-discharge~\cite{C02}, colloidal chemistry~\cite{C11}, sputtering or laser ablation~\cite{C12}, and ripening from gel-like films~\cite{C08}. Each method has its specific advantages and disadvantages. Critical issues are the stability of the particles with time, their protection against oxidation~\cite{C13}, and the control of their size and shape~\cite{C14}. Here we present a novel, simple method to produce inherently protected iron nanoparticles. They are generated by heating iron phthalocyanine crystallites in vacuum. The particles exhibit a narrow diameter distribution and ferromagnetic hysteresis. The obtained structures and particles were imaged by scanning electron microscopy (SEM) and transmission electron microscopy (TEM) after the heat treatment. \textit{In-situ} TEM observations show that the formation process is a complex mechanism taking place basically in two phases, a first classical nucleation, growth and ripening phase and a second phase in which the particles move in the matrix to incorporate material. The second phase shows astonishing similarities with the search for nutrition of living micro-organisms.

\begin{figure}[!h]
\begin{center}
\includegraphics[width=0.45\textwidth]{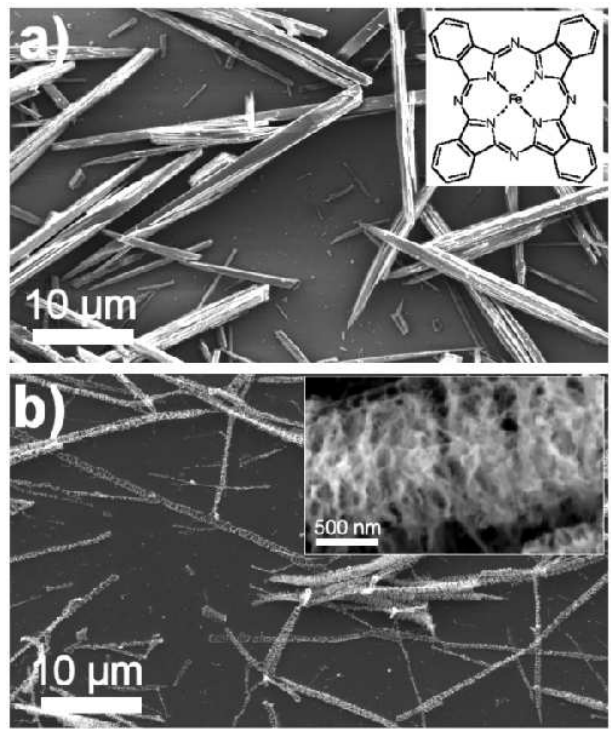}
\caption{\it SEM micrographs of a) as-purchased FePc crystals dispersed on a silicon surface (Inset: Iron phthalocyanine molecule (C$_{32}$H$_{16}$N$_{8}$)Fe), b) the obtained skeleton structure after heating for 30~min at 500$^{\circ}$C at a base pressure of $<2\cdot10^{-5}$~mbar (Inset: Close-up of the transformed crystals).}
\label{F01}
\end{center}
\end{figure}

\begin{figure}[!h]
\begin{center}
\includegraphics[width=0.45\textwidth]{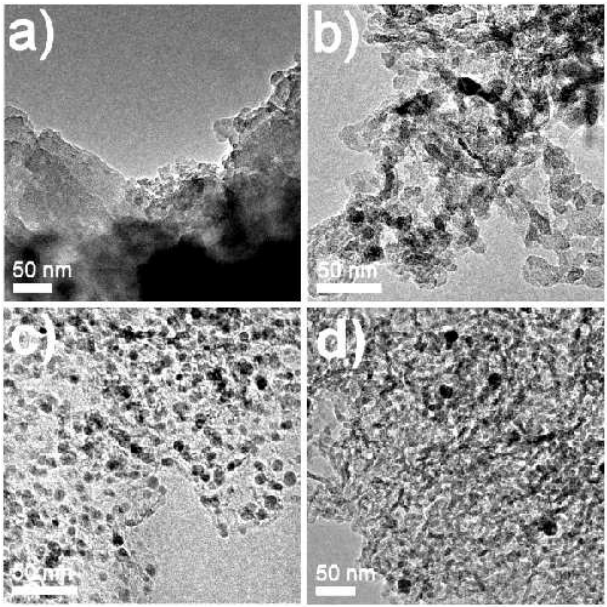}
\caption{\it TEM micrographs of a time series of FePc heated at 500$^{\circ}$C in a vacuum furnace for a) 2~min b) 3~min c) 10~min d) 30~min.}
\label{F02}
\end{center}
\end{figure}

\begin{figure}[!h]
\begin{center}
\includegraphics[width=0.45\textwidth]{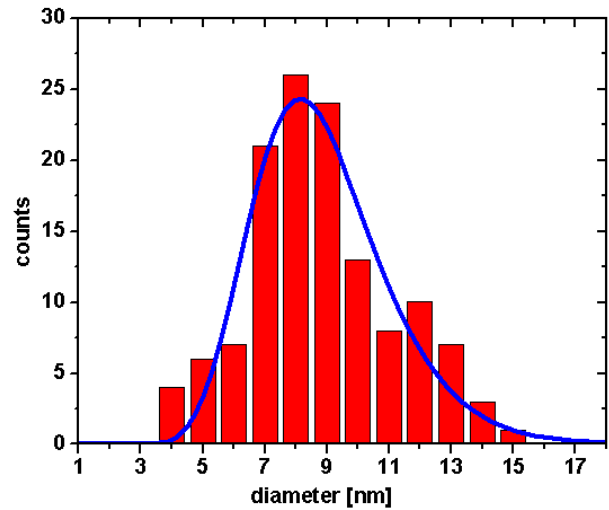}
\caption{\it A statistic analysis of 130 particles after a heat treatment of 10~min at 500$^{\circ}$C (Fig.~\ref{F02}c) allows fitting the distribution of particle diameters with a log-normal distribution. The fitting parameters for the log-normal distribution (blue curve) $y = A \cdot exp[ - (ln(x) - \mu)^{2} / 2\sigma^{2}] / (\sqrt{2\pi} \cdot \sigma \cdot x)$ are the amplitude $A = 123.4$, the standard deviation $\sigma = 0.24$ and mean $\mu = 2.15$. The function peaks at $exp(\mu - \sigma^{2}) = 8.1~nm$ with the standard deviation of $\sqrt{(exp(2\mu + \sigma^{2}) \cdot (exp(\sigma^{2}) - 1))} = 2.15~nm$.}
\label{F03}
\end{center}
\end{figure}

Iron phthalocyanine (C$_{32}$H$_{16}$N$_{8}$)Fe (FePc, Fig.~\ref{F01}a(inset)) as purchased is a black powder consisting of bacillary crystallites. It is a macrocyclic compound having an alternating nitrogen-carbon aromatic ring structure coordinating a central iron atom. Fig. 1a shows such crystals dispersed on a silicon surface out of a dried ethanolic suspension. Such samples were heated in a furnace for 30~min at 500$^{\circ}$C at a base pressure of $<2\cdot10^{-5}$~mbar. During heating the pressure increased due to sublimation of FePc molecules, and evaporation of nitrogen and hydrogen from decomposed FePc molecules. SEM micrographs show that the crystals transform into skeleton-like structures (Fig.~\ref{F01}b). A closer look to the structures reveals a capillary network which probably consists of tubular or flaky carbon (Fig.~\ref{F01}b(inset)). TEM allows the investigation of the internal structure. Heating of FePc crystals on SiO$_{2}$ covered TEM grids in a furnace shows that nanoparticles form in the crystals with elapsing time. Fig.~\ref{F02} demonstrates this transformation: After 2~min heating in a furnace at 500$^{\circ}$C almost no change is observed but already after 3~min the TEM images show cleft structures. After 10~min the nanoparticles become clearly visible and after 30~min they coarsen to bigger clusters. TEM diffraction measurements after the heat treatment show that the nanoclusters consist of fcc-Fe at room temperature.  A statistic analysis of 130 particles after a heat treatment of 10~min at 500$^{\circ}$C (Fig.~\ref{F02}c) reveals a reasonably narrow size distribution peaking at 8.1~nm with a standard deviation of 2.15~nm (Fig.~\ref{F03}).

\begin{figure}[!h]
\begin{center}
\includegraphics[width=0.45\textwidth]{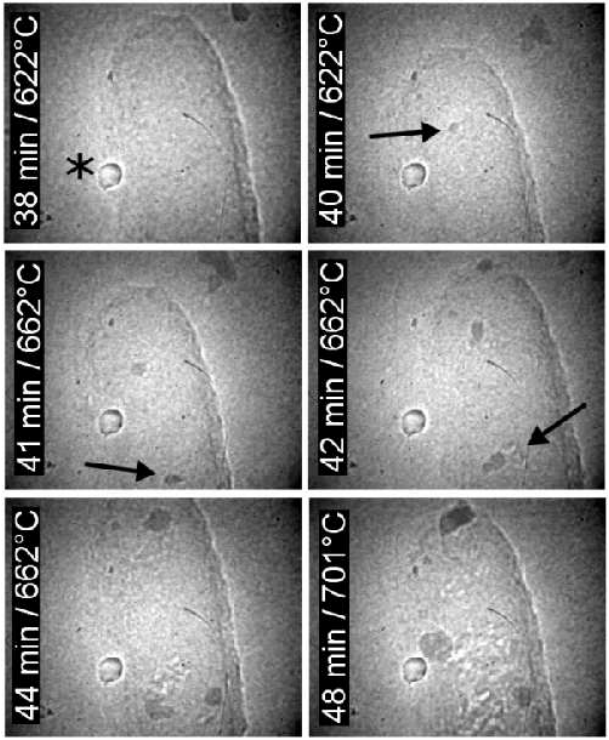}
\caption{\it TEM micrographs of an individual FePc crystal heated \textit{in-situ} up to a temperature of 701$^{\circ}$C. The size of this video cut-out is about 100 $\times$ 80~nm$^{2}$. First, the particles (arrows) grow by a ripening process. Later they start to acquire material by moving in the matrix (* Impurity of the CCD camera}
\label{F04}
\end{center}
\end{figure}

\begin{figure}[!h]
\begin{center}
\includegraphics[width=0.45\textwidth]{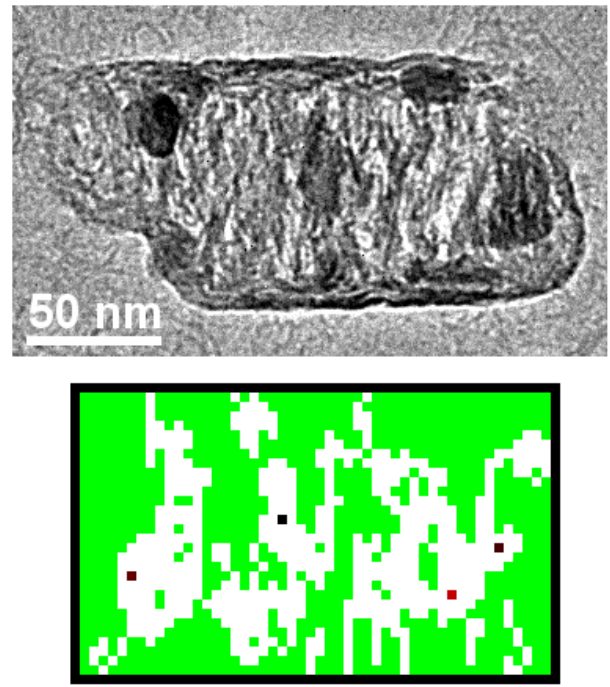}
\caption{\it Comparison of an \textit{in-situ} transformed FePc crystallite and a simulated self-avoiding random walk of pseudo particles with a size of about 3~nm (Green: pristine material; Red-black: pseudo particles with increasing amount of absorbed material).}
\label{F05}
\end{center}
\end{figure}

\textit{In-situ} heating of the samples in a TEM unveils that the transition starts around 400$^{\circ}$C and all individual crystals transform into iron/carbon structures. In this case, the samples were not observed during the whole process but just sporadically during heating. In other experiments, the samples were heated from ambient temperature up to 700$^{\circ}$C under constant observation. In this case, the transition was directly observed between 600 and 700$^{\circ}$C~\cite{C15}. In this temperature range the crystals transform in a few minutes. Fig.~\ref{F04} shows a series of video cut-outs of the development of an individual crystal. In this case the formation of the metal particles starts at 622$^{\circ}$C, then at 662$^{\circ}$C they are getting bigger, and at 701$^{\circ}$C they begin to move in the matrix absorbing material on their way. The nanoparticles move randomly in the Fe/C structure of the original FePc crystal leaving empty space behind. This migration gives rise to the characteristic final structure. The particle movement corresponds to a \textit{self-avoiding random walk} and can be simulated by a simple algorithm: In the simulations the FePc crystals are represented by a 2D matrix. The matrix positions are "filled" with pseudo particles with the size of about 3~nm, the "nutrition" units. Then, a random distribution of germs is spread in the matrix. The migration algorithm assumes that each germ picks up a random neighbor unit and moves to its location, leaving behind a void location where it has been before. In principle, each location is equal as long as there is "food", empty locations are disregarded. Additionally to the random walk we implemented a slight anisotropy in a way that the absorption in vertical direction is favored in order to mimic the crystal structure of FePc. Proceeding with this algorithm for each germ leads to a final structure similar to the transformed FePc crystals. The germs stop their movement when there is no direct neighbouring nutrition unit left. Fig.~\ref{F05} shows a comparison of an \textit{in-situ} transformed FePc crystallite and a simulated transformation.

The remarkable behavior of the FePc crystallites under heating can be explained basically with two distinguished growth phases: First, annealing in the temperature regime between 400$^{\circ}$C and 500$^{\circ}$C leads to a breaking of intramolecular bonds. Hydrogen and nitrogen recombine and evaporate as H$_{2}$ and N$_{2}$. In this initial phase iron is quite mobile and can diffuse since it is atomically distributed in the remaining carbon matrix and weakly bound. Nanoparticles are generated by stationary germs which grow by attracting material from the environment~\cite{C08} and eventually ripening. In a second phase, at an increased temperature of about 650$^{\circ}$C the nanoparticles start to move in the carbon matrix absorbing more material. This behavior results in a separation of iron and carbon. The carbon is reorganised by the moving iron clusters and forms the characteristic final web-like structure. This effect is similar to the movement of biological cells, known as phagokinetic tracks~\cite{C16,C17,C18} and tracks of surface alloy islands~\cite{C19}. Diffusion, ripening, and surface (energy) minimization causes that iron can form bigger roundish aggregates. Additionally, the surface energy can be lowered when the iron aggregates are surrounded by carbon. This leads to a "hiding" of particles in the pockets. The particles must avoid free surface facets (capillary forces). In turn, this leads to a growth of the particles into the material, out of the holes into the self-created cavities. Since the particles stay, by undergoing this process, always in contact with non-transformed material, they can absorb more and more iron (from the stock). The movement stops and leaves the final characteristic structure behind when there is no more iron left in the direct proximity. In the long run particles can diffuse over large distances and form extended aggregates. As larger particles move more slowly, a relatively narrow size distribution can be obtained at intermediate stages. Further optimisation of the process parameters might lead to an even sharper size distribution.

The iron/carbon powder obtained after heating of macroscopic amounts of FePc for 10~min at 500$^{\circ}$C shows ferromagnetic hysteresis. SQUID measurements reveal that both the coercive field and the remanent magnetization increase as the temperature decreases. Furthermore, the iron/carbon structures are active catalysts for the CVD growth of carbon nanotubes. In an experiment using a tube furnace at 650$^{\circ}$C multi-wall nanotubes grow nicely using acetylene as carbon source. They are of similar quality as the CVD tubes obtained using iron-containing solutions as catalyst~\cite{C20}. As other groups showed, carbon nanotubes can also be produced directly by pyrolysis of iron phthalocyanine~\cite{C21,C22,C23}. 

To conclude, under heating in vacuum FePc crystallites decompose into iron nanoparticles embedded in a carbon matrix. By means of in-situ TEM investigations we observed \textit{in-situ} the remarkable formation mechanism of these structures. A classical nucleation, growth and ripening phase is succeeded at higher temperature by a gluttony phase during which the particles move in the carbon matrix, acquiring further material. The synthesis method is very appealing due to its simplicity. It involves only one precursor and only one heating step. The method was demonstrated for the fabrication of iron nanoparticles but can easily be generalized to other systems since there are over 70 ions known which can be accommodated by phthalocyanine~\cite{C24}.

\section*{Experimental methods}

Iron phthalocyanine (FePc) was suspended in ethanol, sonicated for 5~min, and dried on silicon oxide surfaces. The samples were then submitted to a tubular vacuum furnace at 500$^{\circ}$C and $<2\cdot10^{-5}$~mbar for different time periods by means of a transfer rod.

SEM characterizations were performed to analyse the structures in plan view. A Philips XL 30 microscope equipped with a field emission gun operating at an acceleration voltage between 3 and 5~kV, a working distance of typically 10~mm, and in secondary electron image mode was used.

The internal structure of the samples was controlled by TEM. For this purpose a Philips EM 430 microscope equipped with a Gatan image plate operating at 300~kV was used. The FePc crystallites were dispensed on TEM grids covered with a thin SiO2 film. The transformation of FePc crystals as a function of temperature to a carbon matrix with embedded iron particles was directly observed and video-taped \textit{in-situ} in a TEM equipped with a resistively heatable sample holder.

The magnetic properties of the obtained Fe/C powder were measured with a quantum design SQUID magnetometer (MPMS7). This apparatus, which has a sensitivity of $<6\cdot10^{-6}$~emu (corresponding of 1~$\mu$g of powder), is equipped with a cryogenic sample stage which permits control of the temperature between 1.8 and 340~K. The applied field varied between $\pm$45~kOe.

\section*{Acknowledgement}

The Swiss National Science Foundation (SNF) is acknowledged for the financial support. The electron microscopy was performed at the Centre Interdepartmental de Microscopie Electronique (CIME) of EPFL.

\clearpage


\begin{thebibliography}{00}
\bibitem{C15} The difference in transition temperature might be due to a decomposition of organic molecules of the residual pressure by the electron beam and a deposition of amorphous carbon on the samples which inhibits an immediate transformation of the FePc crystals.
\end{thebibliography}
\end{document}